\documentstyle[pra,aps,amsfonts,epsf]{revtex}
\begin{document}
\draft
\twocolumn[\hsize\textwidth\columnwidth\hsize\csname 
@twocolumnfalse\endcsname
\title{A Mach-Zehnder Interferometer for a Two-Photon Wave Packet} 
\author{Luiz Carlos Ryff and P. H. Souto Ribeiro\cite{ca}}
\address{Instituto de F\'{\i}sica, 
Universidade Federal do Rio de 
Janeiro, Caixa Postal 68528, Rio de Janeiro, RJ 22945-970, Brazil} 
\date{\today}
\maketitle
\begin{abstract}
We propose an experiment that permits  observation of the de Broglie
two-photon wave packet behavior for a pair of photons, using a Mach-Zehnder interferometer.
It is based on the use of pulsed lasers to generate pairs of photons via
spontaneous parametric down-conversion and the post-selection of events. It differs
from previous realizations by the use of a third time-correlated photon to engineer
the  state of the photons. The same technique can give
us which-path information via an ``interaction-free'' experiment and can be used in
other experiments on the foundations of quantum mechanics related to
wave-particle duality and to nonlocality.
\end{abstract}
]
\section{Introduction}

Recently, Jacobson {\em et al. }\cite{1} described theoretically an
interferometer capable of measuring the de Broglie wavelength of an ensemble
of photons, which is given by $\lambda $/N, where $\lambda $ is the
wavelength of each photon taken individually, and $N$ is the average number
of constituent photons. In their proposal, they suggest the use of high-Q
cavities. More recently, Fonseca, Monken, and P\'{a}dua \cite{2}, performed
a Young double-slit experiment to demonstrate, in a practical way, how to
measure the de Broglie wavelength of a two-photon wave packet. This
experiment has helped us to understand the role of entanglement in
experiments in which the de Broglie two-photon wave packet behavior is observed
through the frequency of the fourth order interference fringes. Here we wish
to discuss the idea of an experiment which is equivalent to a Mach-Zehnder
(M-Z) interferometer for a two-photon wave packet. 
The scheme produces a two-photon entangled state inside the interferometer.
The behavior of the
two-photon wave packet can be explained in terms of Einstein-Podolsky-Rosen
(EPR) correlations\cite{3}.
A similar idea can be used to create a M-Z
interferometer for three photons. In this case the three photons will be in
a Greenberger-Horne-Zeilinger (GHZ) state\cite{4}.

Interference fringes oscillating with a frequency corresponding to a
two-photon wave packet interference have been observed in previous
experiments. The question of whether or not these doubled frequencies
actually correspond to the interference of a wave packet containing two
photons is more subtle. If one wants to observe two-photon wave
packet interference (TPWI), two main difficulties must be overcome. The
first one is to keep the two photons together, when they are crossing the
interferometer. The second one is to separate the one-photon signal from the
two-photon one in detection. In Ref. \cite{2}, the pair of photons was kept
together by preparing the input state. In Ref. \cite{5}, Martienssen 
{\em et al.} have observed TPWI with a Michelson type interferometer. In
their experiment, the events in which two photons travel together through the
interferometer were selected from the other possibilities. This was
performed by making one of the arms of the interferometer larger than the
other. Thus, whenever each photon followed a different path in the
interferometer, no coincidence was registered. In both experiments, the
two-photon detection was performed by photon coincidence. A similar approach
was used by Burlakov {\em et al.} \cite{6}, where state preparation was
employed in the observation of the TPWI.

Our proposal is based on the technique of generation of photon pairs via
spontaneous parametric down-conversion (SPDC) using pulsed pump lasers \cite
{7}. The use of pulsed pumping allows us to have a temporal correlation
between two initially independent photons, thereby engineering entangled states
with more than two photons. Our approach can be used to investigate
questions related to the foundations of Quantum Mechanics, such as
wave-particle duality and nonlocality, for example.

\section{Utilizing Three Correlated Photons To Observe Two-Photon Wave
Packet Interference}

\begin{figure}[h]
\vspace*{5cm}
\epsfxsize=8cm
\epsfysize=4cm
\epsfbox{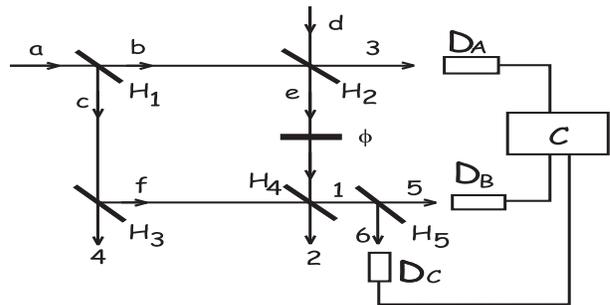}
\caption{Sketch of a M-Z interferometer for observing Two-Photon
Wave Packet Interference.}
\label{fig1}
\end{figure}

The experiment we are proposing is represented in Fig.\ref{fig1}. Two
photons with equal polarizations, which can be generated via type-I SPDC,
follow path $a$ and arrive together at beam-splitter (B.S.) H$_1$. The M-Z
interferometer is built with four 50-50 beam splitters (H$_1$,H$_2$,H$_3$,
and H$_4$). There are three possibilities for the paths followed by the
photons: (1) one photon follows path $b$ and the other follows path $c$, (2)
both photons follow path $b$, and (3) both follow path $c$. To discard the
first possibility, a third photon, with the same polarization as the other
two, is introduced via path $d$. The three temporally correlated photons can
be generated using pulsed lasers \cite{7}, so that whenever one or two
photons follow path $b$ they arrive at H$_2$ together with the photon
following path $d$. From the different events which can take place, we are
only interested in those in which only one photon is registered by detector D$_{A}$. 
(To be sure that only one photon has been registered by detector D$_{A}$, we
use a triple coincidence circuit, as represented in Fig.\ref{fig1}.) In this
case, there are two possibilities: (1) No photon follows path $e$, which
implies that two photons follow path $c$, or (2) two photons follow path $e$.
If only one photon follows path $b$, after H$_2$ we will have two photons
following the same direction \cite{8,9}, that is, both will either impinge
on detector D$_{A}$ or follow path $e$, and no triple coincidence will be
registered. To complete the interferometer for two photons, we introduce B.
S. H$_5$ and use a triple coincidence circuit, so that only events in which
two photons emerge together from the interferometer will be considered.
In these circumstances, the gated two-photon detection rate R$_{11}$ at site $1$
is proportional to $1-\cos 2\phi $. (The triple coincidence detection rate at sites
3,5 and 6 allows us to observe this dependence on 2$\phi$, since it is proportional to
R$_{11}$.) To see this, let us write the symmetrized three-photon initial state as \cite{10} 
\begin{equation}
N(|a\rangle |a\rangle |d\rangle +|a\rangle |d\rangle |a\rangle +|d\rangle
|a\rangle |a\rangle ),  \label{eq1}
\end{equation}
where $N$ stands for the normalization factor and $|a\rangle $ ($|d\rangle $%
) represents a photon coming via path $a$ ($d$). The action of the B. S. s
can be represented as $|a\rangle \rightarrow (1/\sqrt{2})(|b\rangle \\%
+i|c\rangle )$, and so on. It is then easy to show that the three-photon
state, after H$_2$ and H$_3$, is given by 
\begin{equation}
N\left[ \left( |f\rangle |f\rangle +|e\rangle |e\rangle \right) |3\rangle
+symmetrization\,terms\right] ,  \label{eq2}
\end{equation}
where only the terms which in principle can lead to the triple coincidence
detection we are interested in  have been taken into account. (Please note
that $|3\rangle $ represents the state of one photon following
direction $3$, it is {\em not} a three-photon Fock state.) The action of the phase
shifter can be represented as $|e\rangle \rightarrow e^{i\phi }|e\rangle $.
Hence, if a photon is detected at site $3$, the state $(2)$ is reduced to
the two-photon state 
\begin{equation}
N(|f\rangle |f\rangle +e^{2i\phi }|e\rangle |e\rangle ).  \label{eq3}
\end{equation}
From $(3)$ we then obtain 
\[
N\left[ (1-e^{2i\phi })|1\rangle |1\rangle +i(1+e^{2i\phi })|1\rangle
|2\rangle \right. 
\]
\begin{equation}
\left. +i(1+e^{2i\phi })|2\rangle |1\rangle -(1-e^{2i\phi })|2\rangle
|2\rangle \right] ,  \label{eq4}
\end{equation}
for the two-photon state after H$_4$. From $(4)$ we obtain the gated(or triple) coincidence
detection rates: 
\begin{equation}
R_{11}=R_{22}\propto P_{11}=P_{22}\propto \left| 1-e^{2i\phi }\right|
^2\propto 1-\cos 2\phi  \label{eq5}
\end{equation}
and 
\begin{equation}
R_{12}=R_{21}\propto P_{12}=P_{21}\propto \left| 1+e^{2i\phi }\right|
^2\propto 1+\cos 2\phi ,  \label{eq6}
\end{equation}
where the phase shift is given by $\phi =2\pi \delta l/\lambda $, in which $
\delta l$ is a small difference between the two path lengths in the
interferometer and $\lambda $ is the wavelength of each photon. We see that
in some respects the two photons behave as a single particle of wavelength $
\lambda /2$. But when $\phi =0$ and R$_{11}=0$, this does not mean that the
two photons are following direction $2$. Actually, as we see from $(5)$ and $
(6)$ each photon follows a different direction, that is, one photon goes to
channel $1$ and the other to channel $2$.

\section{Fringes With Doubled Frequency But Without Two-Photon Wave
Packet Interference}

\begin{figure}[h]
\vspace*{5cm}
\epsfxsize=8cm
\epsfysize=4cm
\epsfbox{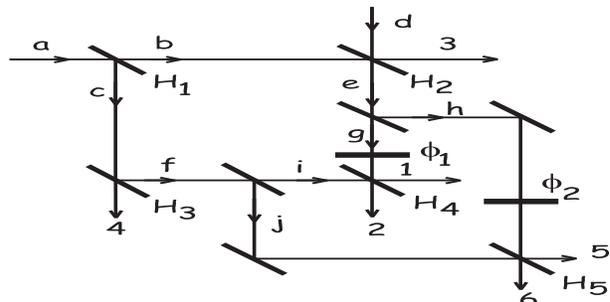}
\caption{An extension of the experiment represented in Fig. 1, to disclose its 
nonlocal features.}
\label{fig2}
\end{figure}

Although in agreement with de Broglie's relation, $\lambda =h/p$, it may
sound a bit surprising that two photons when moving together have a
wavelength which is half the wavelength of each photon taken individually,
since there is no interaction keeping the photons together.  
The physical property connecting the photons is entanglement. 
The pair of photons is not in an entangled state initially, at the interferometer
entrance. The entanglement is induced by the third photon. In order to provide
a clearer understanding of this aspect, let us consider
a modified version of the
experiment represented in Fig. \ref{fig1}, as represented in Fig. \ref{fig2}. 
Now we are interested in the situation in which one photon is detected at 
$3$ and the other two impinge on H$_4$ and H$_5$, respectively. From the
previous discussion, we see that the third photon coming via path $d$ acts
as a sort of ``catalyst'', inducing entanglement between the other two
photons (actually, a photon has no individuality, and it makes no sense to
say that the photon at site $3$ is the {\it same }photon which came via path 
$d$, so it is an abuse of language to refer to the ``other two photons'').
Since, in our experiment, the two photons always follow the same path in the
first interferometer, if one photon impinges on H$_4$ via path $g$ ($i$),
the other can only impinge on H$_5$ following path $h$ ($j$). That is, we
have a direction-entangled two-photon state. As is well known \cite
{11,12,13,14,15,16,17}, this leads to the phenomenon of two photon
interference and to the violation of a Bell inequality, a signature of
quantum nonlocality \cite{18}. To see this, let us consider in Fig. \ref
{fig2} the situation in which a photon has been detected at $3$ and the
remaining two-photon system has already been reflected at H$_2$ and H$_3$
but has not yet reached the next B. S.s . From $(2)$ and $(3)$, we see that
the two-photon state is given by 
\begin{equation}
N(|f\rangle |f\rangle +|e\rangle |e\rangle ).  \label{eq7}
\end{equation}
It is then easy to see that after the phase shifters the two-photon state is 
\begin{equation}
N\left[ |i\rangle |j\rangle +|j\rangle |i\rangle +e^{i(\phi _1+\phi
_2)}|g\rangle |h\rangle +e^{i(\phi _1+\phi _2)}|h\rangle |g\rangle \right] ,
\label{eq8}
\end{equation}
where the terms in which the two photons follow the same path together are
not included, since we are not interested in these events. Thus, after H$_4$
and H$_5$ the two-photon state will be 
\begin{eqnarray}
&&N\biggr\{ \left[ 1-e^{i(\phi _1+\phi _2)}\right] (|1\rangle |5\rangle
-|2\rangle |6\rangle )+i\left[ 1+e^{i(\phi _1+\phi _2)}\right] \\
&&(|1\rangle |6\rangle +|2\rangle |5\rangle )\biggr\} ,  \nonumber
\label{eq9}
\end{eqnarray}
where the symmetrization terms have not been included, since they add no new
information to our discussion. Comparing the expressions $(4)$ and $(9)$ we
see that the experiment represented in Fig. \ref{fig1} is a degenerate case
of the experiment represented in Fig. \ref{fig2}, which is similar to the
experiments discussed in $[11]$ and $[17]$. Therefore, the behavior of the
two photons as a single entity can be understood as a consequence of the
nonlocal correlations of the entangled photons. 
The interference fringes depend on $\phi _1+\phi _2$. 
When $\phi _1$ = $\phi _2$ = $\phi $, they will oscillate
with 2$\phi $ as in the TPWI. However, each photon impinges on a different
B. S..

\section{Fringes With Doubled Frequency For Separated Photons}

The previous discussion raises an interesting point, which shows how the
photon picture can sometimes be misleading  \cite{19}. For example, let us
consider the experiment represented in Fig. \ref{fig1}, but without the
third photon coming via path $d$ and with H$_{5}$ removed.
We are only interested in the
situation in which the two photons coming via path $a$ arrive at H$_4$.
Therefore, H$_{2}$ and H$_{3}$ now represent perfect mirrors.
This is an experiment in which Dirac's
dictum ``each photon interferes only with itself'' \cite{20} must apply.
That is, if a light beam is sent to the interferometer via path $a$, the
intensity observed at sites $1$ and $2$ will only depend on the phase $\phi $, 
not on the mean number of photons in the beam. However, in our experiment
we might be tempted to assume that we either have each photon following a
different path or we have both photons following the same path together. In
the first case, they would arrive together at H$_4$ and follow together the
same direction \cite{8,9}, which could,  with equal probability, be either 
$1$ or $2$. In the second, their behavior would depend on $2\phi $, as we
have just seen. However, the interesting thing is that, although these two
different behaviors can be discerned from the quantum mechanical formalism,
the final result, given by the probabilities of detection, is the same as we
would observe by considering the photons as totally independent. That is, 
we would observe the same result if we
first sent a photon and detected it, and only then sent and detected the second
photon. This can be seen as follows. After H$_2$, H$_3$, and $\phi $, 
the initial two-photon state $|a\rangle |a\rangle $ becomes 
\begin{equation}
\frac 12\left( -e^{2i\phi }|e\rangle |e\rangle -ie^{i\phi }|e\rangle
|f\rangle -ie^{i\phi }|f\rangle |e\rangle +|f\rangle |f\rangle \right) .
\label{eq10}
\end{equation}
Hence, after H$_4$ we obtain 
\[
\frac 14\left[ \left( 1-2e^{i\phi }+e^{2i\phi }\right) |2\rangle |2\rangle
-i\left( 1-e^{2i\phi }\right) |2\rangle |1\rangle \right. 
\]
\begin{equation}
\left. -i\left( 1-e^{2i\phi }\right) |1\rangle |2\rangle -\left( 1+2e^{i\phi
}+e^{2i\phi }\right) |1\rangle |1\rangle \right]  \label{eq11}
\end{equation}
for the two-photon state. From $(11)$ we see that the probability amplitude 
for having the two photons being detected in the same channel depends on $2\phi $ (which
corresponds to both photons following the same path together) and on 
$\phi $ (which corresponds to each photon following a different path). On the other
hand, the probability amplitude 
for having  each photon being detected in a different
channel depends only on $2\phi $, since whenever they follow different paths
they emerge from H$_4$ in the same
channel. Nevertheless, the detection probabilities are exactly the same as
we would have if instead of having the two photons arriving together at the
interferometer we had them arriving at totally independent and different
times. To see this, let us calculate the detection probabilities in an ideal
situation. If only one photon is sent to the interferometer, the detection
probability amplitudes are given by

\begin{eqnarray}
A_1 &=&\frac 12(1+e^{i\phi })  \label{eq12} \,\,\mbox{and }\\
A_2 &=&\frac 12(1-e^{i\phi }).  \nonumber
\end{eqnarray}
Thus, the probabilities are given by

\begin{eqnarray}
P_1 &=&\left| A_1\right| ^2=\frac 14(1+e^{-i\phi })(1+e^{i\phi })
\label{eq13} \,\,\mbox{and}\\
P_2 &=&\left| A_2\right| ^2=\frac 14(1-e^{-i\phi })(1-e^{i\phi}).  \nonumber
\end{eqnarray}
Considering the photons as independent, the probabilities of ``coincident''
detections (in this case, we have used the term``coincidence '' for detections
of individual photons at the same output port,  P$_{11}$ and P$_{22}$,
or at diferent output ports, P$_{12}$ and P$_{21}$) are given by 
\begin{eqnarray}
P_{11} &=&P_1^2=\frac 1{16}(1+e^{-i\phi })^2(1+e^{i\phi })^2,  \label{eq14}
\\
P_{12} &=&P_{21}=P_1P_2=\frac 1{16}(1-e^{-2i\phi })(1-e^{2i\phi })  \nonumber
\,\,\mbox{and}\\
P_{22} &=&P_2^2=\frac 1{16}(1-e^{-i\phi })^2(1-e^{i\phi })^2. \nonumber
\end{eqnarray}
On the other hand, using $(11)$ we see that 
\begin{eqnarray}
A_{11} &=&\frac 14(1+e^{i\phi })^2,  \label{eq15} \\
A_{12} &=&A_{21}=\frac 14(1+e^{i\phi })(1-e^{i\phi })  \,\, \mbox{and}\nonumber \\
A_{22} &=&\frac 14(1-e^{i\phi })^2.  \nonumber
\end{eqnarray}
Hence, 
\begin{eqnarray}
P_{11} &=&\left| A_{11}\right| ^2=P_1^2  \label{eq16} \\
P_{12} &=&P_{21}=\left| A_{12}\right| ^2=P_1P_2  \,\,\mbox{and}\nonumber \\
P_{22} &=&\left| A_{22}\right| ^2=P_2^2.  \nonumber
\end{eqnarray}
We obtain the same result.

From $(11)$, we see that interference patterns with oscillations depending
on 2$\phi $ can be observed by performing coincidence measurements between
output channels 1 and 2. Because the photons split at the output of the
interferometer, it would be questionable to interpret this result as a TPWI.
Moreover, from $(16)$, we see that we can also have oscillating patterns
depending on 2$\phi $ if we send one photon at a time through the
interferometer. This would be even harder to interpret as a TPWI.

For the three-photon version of the M-Z interferometer we have just
presented, the interference patterns depending on 2$\phi $ are obtained, and
the interpretation in terms of TPWI is always possible.

\section{Three-Photon Wave Packet Interference}

\begin{figure}[h]
\vspace*{5cm}
\epsfxsize=8cm
\epsfysize=4cm
\epsfbox{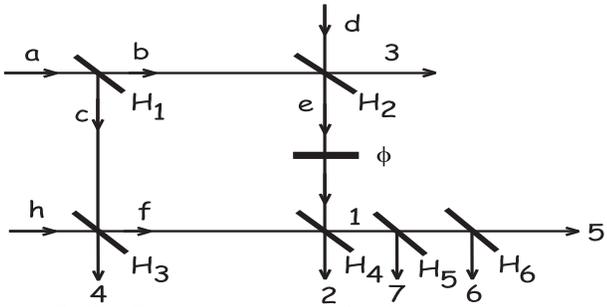}
\caption{Sketch of a M-Z inteferometer for observing Three-Photon Wave Packet
Inteference.}
\label{fig3}
\end{figure}

The same idea can be extended to build an interferometer for a three-photon
wave packet. In this case, three photons impinge on H$_1$, a fourth photon
impinges on H$_2$, a fifth photon impinges on H$_3$, and we register the
fivefold coincidence detection at sites $3$, $4$, $5$, $6$, and $7$ (Fig. 
\ref{fig3}). It is easy to see that now we have three photons following the
same path together. If only one photon is detected at site $3$, it is
possible to infer that either no photon, two photons, or three photons
follow path $e$. Similarly, if only one photon is detected at site $4$,
either no photon, two photons, or three photons follow path $f$. Since three
photons impinge on H$_4$, the only possibilities are no photon following
path $e$ and three photons following path $f$, or three photons following
path $e$ and no photon following path $f$. Moreover, the symmetry of the
situation allows us to conclude that the probabilities for these two
possibilities are the same. Now we have a $3\phi $ dependence in the
interference. It is easy to see that the two photons coming via paths $d$
and $h$ induce entanglement between the three photons impinging on H$_4$,
generating a GHZ state.

\section{Discussion}

In this paper we have proposed a way to achieve, at least for the cases of
two and three photons, what Yamamoto has imagined, that is, a B.S. that
splits a many-photon wave packet as a whole. We have also shown that the
many-photon wave packet behavior is a consequence of entanglement. From this
point of view, we see that there is a much simpler way to entangle the
photons in the interferometer: instead of having the two photons coming via
path $a$, we can have one photon coming via path $a$ and the other coming
perpendicularly to path $a$. After H$_1$ they will necessarily follow the
same direction. However, this procedure cannot be extended to the situation
in which we have a three-photon wave packet, and it is not a realization of
Yamamoto's idea. We have also  seen that the assumption according to which
the events are ``out there'' and we are selecting only those in which the
two photons go together may be misleading. Naturally, our explanation of the
many-photon wave packet behavior in terms of EPR correlations can in
principle be extended to any system of particles, including atoms and
molecules, for example, since to the extent that the particles have to go
together, either along one path or the other, after a B. S., they are in a
direction-entangled state.

We would like to point out that the introduction of an additional photon via
path $d$, as we are proposing (Fig. \ref{fig1}), can give us
``interaction-free'' which-path information. For example, if we have only
one photon coming via path $a$ and another coming via path $d$, whenever
only one photon is detected at site $3$ it is possible to infer that the
photon coming via path $a$ has followed path $c$. In this way, it is
possible to obtain information about the path followed by the first photon
without directly interacting with it. This suggests interesting new 
experiments on the foundations of quantum mechanics, more specifically, on
wave-particle duality and pilot wave interpretation, as well as on
quantum nonlocality, to be discussed in a paper to follow.

\section{Conclusion}

A new version of a Mach-Zehnder interferometer is presented. Two temporally
correlated photons are sent through the input port, while a third one is
sent through another port. As a result, interference patterns oscillating
with a frequency two times larger than that of a single photon are observed
when triple coincidence is detected. We show that they can be interpreted as
two-photon wave packet interference. The same kind of pattern can be
observed for different experimental arrangements, but in most of them the
interpretation in terms of two-photon wave packet interference is not
possible. Another version of the M-Z interferometer for three-photon wave
packet interference is also presented. The schemes can be useful for
studying wave-particle duality and quantum nonlocality.

\section{ACKNOWLEDGMENT}

Financial support was provided by Brazilian agencies CNPq, PRONEX, FAPERJ
and FUJB.

\vspace*{1cm}

\begin{references}
\bibitem{18}  As in Franson's experiment \cite{17}, in our experiment we
obtain pairs of photons entangled in direction via the post-selection of
events. In this respect, the two experiments are similar. But in Franson's
experiment we need unbalanced interferometers and we cannot use pulsed
lasers, since we must have uncertainty about the time of emission.
\end{references}
\end{document}